# Fermionic Approach to Elementary Excitations and Magnetization Plateaus in an $S = 1/2$ XX Hybrid Trimer–Dimer Chain

K. S. Chikara, A. K. Bera*, A. Kumar*, S. M. Yusuf *

*Solid State Physics Division, Bhabha Atomic Research Centre, Mumbai, 400085, India
and Homi Bhabha National Institute, Anushaktinagar, Mumbai 400094, India*

* Contact Authors: akbera@barc.gov.in, amitkr@barc.gov.in, smyusuf@barc.gov.in

**Abstract:**

We study the elementary excitations and magnetization of a one-dimensional spin-1/2 XX chain comprising trimer-dimer units ($J_1$-$J_1$-$J_2$-$J_3$-$J_2$ topology) under a transverse magnetic field ($h$). Using Green's function theory and the Jordan-Wigner transformation, we map the system to spinless fermions and focus on antiferromagnetic (AFM) interactions. At zero temperature, distinct 1/5 and 3/5 magnetization plateaus emerge, determined by the overall periodicity $Q = 5$, with the number of plateaus matching the number of excitation gaps above the Fermi level of spinless fermions. The magnetic phase diagram in the ($h$-$Js$) plane features a Luttinger liquid (LL) state, a gapless AFM state, two magnetization plateau state, and a fully polarized gapped magnetic state. The widths of the Luttinger liquid (LL) and gapless AFM phases are found to be proportional to the bandwidths ($\gamma \propto |E(k = 0) - E(k = \pi)|$) of the respective elementary excitations. Whereas, the widths of the magnetization plateau states are found to be linked to the gaps between excitations. Our study opens a new direction for exploring interacting trimer-dimer spin chains in quantum magnetism using experimental techniques, such as neutron scattering and/or through other theoretical/numerical approaches, including Quantum Monte Carlo (QMC) and Density Matrix Renormalization Group (DMRG) methods. Furthermore, our study extends the Oshikawa–Yamanaka–Affleck (OYA) condition to generalized cluster chains, showing that the allowed magnetization plateau states are governed by the global periodicity of the chain (e.g., $Q = 5$ for a trimer–dimer chain), rather than the local periodicity of individual units ($Q = 3$ for a trimer or $Q = 2$ for a dimer).

## I. INTRODUCTION:

The study of quantum phases of matter, particularly in low-dimensional systems where strong quantum fluctuations play a crucial role, has yielded a range of exotic properties [1]. Quantum fluctuations give rise to novel quantum mechanical phenomena such as the fractional quantum Hall effect [2], chiral spin liquids [3], field-induced fractional magnetization plateau states [4,5], and novel quasiparticle excitations [6] that cannot be accounted by classical theories. In this respect, one-dimensional spin systems are of special interest. For instance, gapped magnon excitations have been observed in one-dimensional Heisenberg antiferromagnets with integer spin quantum numbers, in contrast to their spin-1/2 counterparts, which exhibit gapless fractional spinon excitations [7-9].

In particular, an intriguing phenomenon in low-dimensional spin systems is the appearance of quantum magnetization plateaus under the applied magnetic field [10-12]. The plateaus occur when the magnetization quenches to specific value and remain constant despite the variation of external magnetic field. This phenomenon was first observed in various one-dimensional antiferromagnetic (AFM) spin models [13,14], where the magnetization per site exhibits plateaus at rational values of the saturation magnetization. The quantum magnetization plateaus emerge as hallmark features of quantized spin excitations with energy gaps and get stabilized in a spin chain system by topology of the periodic exchange interactions. In contrast, classical magnetization plateaus are associated with field induced spin flip transition of ordered antiferromagnetic states [15]. For example, in a triangular lattice, a 1/3 magnetization plateau is observed for the up–up–down (UUD) ordered spin configuration. In contrast, for a spin-cluster system, viz., a spin-trimer system, the magnetization plateau arises due to quantum entanglement within the ground state. Here, the occurrence of such quantum magnetization plateaus is strongly influenced by the topology of the periodic exchange interactions, spin value, and the relative strength of competing exchange interactions. Theoretical progress over the past two decades has significantly advanced our understanding of magnetization plateaus [16]. In particular, Oshikawa, Yamanaka, and Affleck [17] proposed a general criterion for the existence of magnetization plateaus in quantum spin chains, drawing on arguments similar to those of Lieb, Schultz, and Mattis [18,19], as well as heuristic reasoning based on bosonization approaches. According to their criterion, plateaus can occur only at specific fractional values of saturation magnetization $\frac{M}{M_s}$ that satisfy the following quantization condition:

$$QS\left(1 - \frac{M}{M_s}\right) \in Z \quad (1)$$

where $Q$ is the number of magnetic ions in the elementary cell or in the ground state spin cluster (defining the topology of the periodic exchange interactions), $S$ is the spin quantum number, and $Z$ is an integer.

This condition is necessary, but not sufficient. While a plateau must satisfy Eq. (1), not every allowed fraction yields a stable plateau. Extensive numerical and analytical works support this criterion. For example, by Density Matrix Renormalization Group (DMRG) theory for trimer spin chains ($S = 1/2$, 1, and 3/2) Gu et al. [20] [21] demonstrated that magnetization plateaus occur only for dominant intratrimer and/or intertrimer antiferromagnetic (AFM) interactions. Experimentally, a 1/3 magnetization plateau has been reported for spin-1/2 coupled trimer chain compounds $Na_2Cu_3Ge_4O_{12}$ [22] and $SrMn_3P_4O_{14}$ [23]. Beyond spin trimer chains, related studies have been extended to polymerized spin chains [24], spin ladders [25-27], layered spin systems [28-30], and mixed spin chains [31]. As evident from the literature, magnetization plateaus originate from the topological periodic structure of the ground state, typically found in spin clusters such as dimers [29] ($Q$ =2), trimers [16] ($Q$ =3), tetramers [32,33] ($Q$ =4), or diamond chains [34] ($Q$ =4). Most reported studies were focused on the spin-systems composed of a single type of cluster, such as dimer or timer or tetramer. However, the behaviour of magnetization plateaus in spin-chains composed of hybrid clusters such as dimer–trimer, dimer–tetramer, or trimer–tetramer remains unexplored. The hybrid cluster-spin-chains may offer a promising avenue for uncovering new physics in quantum magnetism. A fundamental question is whether the physical properties arise from one of the dominant spin clusters, or from both of them, or from the global periodicity.

Here, we have carried out numerical study on the hybrid trimer-dimer spin-1/2 chain to elucidate the nature of magnetization plateaus, as well as shed light on their microscopic origin from low-energy excitations. For a hybrid trimer-dimer spin-1/2 chain, magnetization plateau states are theoretically expected at 1/3 or/and 1/5, or/and 3/5 of $M_s$. Interestingly, real materials such as $Na_2Cu_5Si_4O_{14}$ and $Li_2Cu_5Si_4O_{14}$ [35-37] [38] [39] show hybrid trimer-dimer spin chains of $Cu^{2+}$ ions ($S$ =1/2). In the present numerical work, we explore the role of intercluster exchange interactions in a model hybrid trimer-dimer spin-1/2 using the Jordan-Wigner (JW) transformation. The JW transformation, a foundational tool in theoretical physics for nearly a century, maps spin-1/2 operators onto spinless fermion operators, thereby connecting spin dynamics with fermionic behavior [40-42]. This transformation was originally developed for one-dimensional systems [43], and then has been extended to higher dimensional lattices and systems with $S$ >1/2 [44-47]. In all these exact mappings, the spin degrees of freedom are ultimately represented by canonical spinless fermion operators [48]. By applying the JW transformation to a quantum hybrid trimer-dimer spin-1/2 XX chain, we aim to gain deeper insights into the mechanisms responsible for the formation of magnetization plateaus states and to characterize the nature of fermionic elementary excitations. Through a combination of analytical and numerical approaches, we determine the phase diagram in the $h$-$J$ plane and shed light on the underlying physics of interacting trimer-dimer spin-chain systems. Our findings establish a generalized condition for magnetization plateaus in hybrid trimer-dimer spin-1/2 chains, extending beyond the conventional OYA framework. We established that the global periodicity dictates the allowed plateau states, rather than the local periodicity of individual clusters. Moreover, we reveal a rich variety of quantum phases—including the Luttinger liquid (LL) state, gapless AFM states, and magnetization plateau states as well as their intrinsic excitations. Further, we demonstrate that both the stability and width of the plateau states are governed by the excitation spectra, providing new insights into the interplay between exchange interactions in hybrid trimer–dimer systems. In addition, these results are expected to encourage future theoretical and experimental investigations employing complementary techniques such as neutron scattering, Quantum Monte Carlo, and Density Matrix Renormalization Group (DMRG) theory to comprehensively examine the magnetic, thermodynamic, and quasiparticle excitation properties of such hybrid cluster spin systems.

The structure of this paper is as follow. In the next section, we present the model Hamiltonian for the interacting spin-1/2 XX trimer- dimer chain. The topology of the exchange interactions is described by the sequence $J_1$-$J_1$-$J_2$-$J_3$-$J_2$, where $J_1$ represents the intratrimer, $J_2$ the inter trimer-dimer, and $J_3$ the intradimer exchange interactions. By employing Green's function theory in conjunction with the Jordan-Wigner transformation, we compute the ground state magnetic properties of the system. Subsequently, we calculate the magnetization curves for various combinations of exchange interaction strengths $J_1$, $J_2$, and $J_3$, and discuss the emergence and characteristics of quantum magnetization plateaus, as well as the nature of elementary excitations. We also analyze the behavior of magnetization as a function of external magnetic field at finite temperatures, with particular attention to the temperature-induced melting of magnetization plateaus.

## II.  MODEL AND METHOD

### 1.1. Model Hamiltonian

A spin-1/2 chain having hybrid cluster of interacting trimer-dimer units under an external magnetic field is described by the Hamiltonian [Eq. (2)]. A schematic representation of such spin system is illustrated in Fig. 1.

$$H = \sum_{l=1}^{N} \left\{ \left[ J_1(S_{a,l} \cdot S_{b,l} + S_{b,l} \cdot S_{c,l}) + J_2\{S_{c,l} \cdot S_{d,l} + \frac{1}{2}(S_{e,l-1} \cdot S_{a,l} + S_{e,l} \cdot S_{a,l+1})\} + J_3(S_{d,l} \cdot S_{e,l}) \right] - h(S_{a,l}^z + S_{b,l}^z + S_{c,l}^z + S_{d,l}^z + S_{e,l}^z) \right\} \quad (2)$$

Here, $N$ is the numbers of clusters (trimer-dimer) units in the given chain, $h$ is the magnetic field along the transverse ($z$) direction. For simplicity, we set $g\mu_B$=1. The index $l$ identifies the unit cell number, $S$ represents the spin=1/2 operators, and $J_i$ denotes the exchange interactions. We primarily consider an XX-type spin system with antiferromagnetic couplings. The spin trimer is characterized by an intratrimer exchange interaction $J_1$, while the spin dimer has an intradimer exchange interaction $J_3$. The trimer and dimer units are coupled via an inter trimer-dimer exchange interaction $J_2$. Accordingly, the model corresponds to a one-dimensional isotropic XX spin-1/2 system, where each repeating unit in $l^{th}$ unit cell follows a cluster topology defined by sequence $J_1$-$J_1$-$J_2$-$J_3$. Each $l^{th}$ unit interacts with its neighboring ($l$-1)$^{th}$ and ($l$+1)$^{th}$ units via the exchange interaction $J_2$. The spin interactions in Eq. (2) can be conveniently expressed using the spin raising and lowering operators, defined as $S^{\pm} = S_x \pm iS_y$.

$$S_{\alpha,\delta} \cdot S_{\beta,\delta'} = \frac{1}{2}[S_{\alpha,\delta}^+ \cdot S_{\beta,\delta'}^- + S_{\beta,\delta'}^+ \cdot S_{\alpha,\delta}^-] \quad (3)$$

where $\alpha$ and $\beta$ = $a$, $b$, $c$, $d$ and $e$, and $\delta$ and $\delta'$ = $l$, $l$-1, $l$+1 with $l$ = 1…$N$. Therefore, the model Hamiltonian for XX spin-1/2 of interacting trimer-dimer chains can be written as

$$H = \frac{1}{2}\sum_{l=1}^{N} \left\{ \left[ J_1(S_{a,l}^+ S_{b,l}^- + S_{b,l}^+ S_{c,l}^-) + J_2\{S_{c,l}^+ S_{d,l}^- + \frac{1}{2}(S_{e,l-1}^+ S_{a,l}^- + S_{e,l}^+ S_{a,l+1}^-)\} + J_3(S_{d,l}^+ S_{e,l}^-) \right] + h.c. - 2h(S_{a,l}^z + S_{b,l}^z + S_{c,l}^z + S_{d,l}^z + S_{e,l}^z) \right\} \quad (4)$$

where, "*h.c.*" denotes the Hermitian conjugate. In order to compute the elementary excitation spectrum, we employed the green function approach along with JW transformation.

Jordan–Wigner representation of spins utilizes the fundamental feature of a non-interacting gas of fermions. A non-interacting gas of fermions is inherently highly correlated due to the Pauli exclusion principle, which imposes a hard-core constraint that prevents multiple fermions from occupying the same quantum state. Classically, a spin is represented by a vector pointing in a specific direction. Such a representation is applicable for quantum spins with extremely large value of *S*, but once the value of spin *S* becomes small, the behavior of spin changes significantly. In such cases, spins exhibit inherently quantum characteristics, including zero-point motion, and the excitation spectrum becomes discrete, as observed experimentally in various spin cluster systems [6,46]. In general, it is challenging to determine the physics of many-body system with quantum spins, because spin do not conform to the conventional behavior of either fermions or bosons. Nevertheless, spin-1/2 can effectively be treated as fermions in a one-dimensional spin chain system. In a seminal contribution, Jordan and Wigner proposed that the "down" and "up" states of a single spin can be treated as empty and occupied fermionic states. Therefore, the Jordan-Wigner transformation can be unambiguously defined for a linear chain, where all sites can be subsequently enumerated. This formalism has been extensively employed in the literature [34]. For the present case, we have

$$S_{a,l}^+ = a_l^\dagger \prod_{j<l}(1 - 2a_j^\dagger a_j)(1 - 2b_j^\dagger b_j)(1 - 2c_j^\dagger c_j)(1 - 2d_j^\dagger d_j)(1 - 2e_j^\dagger e_j)$$

$$S_{b,l}^+ = b_l^\dagger \prod_{j<l+1}(1 - 2a_j^\dagger a_j) \prod_{j<l}(1 - 2b_j^\dagger b_j)(1 - 2c_j^\dagger c_j)(1 - 2d_j^\dagger d_j)(1 - 2e_j^\dagger e_j)$$

$$S_{c,l}^+ = c_l^\dagger \prod_{j<l+1}(1 - 2a_j^\dagger a_j)(1 - 2b_j^\dagger b_j) \prod_{j<l}(1 - 2c_j^\dagger c_j)(1 - 2d_j^\dagger d_j)(1 - 2e_j^\dagger e_j)$$

$$S_{d,l}^+ = d_l^\dagger \prod_{j<l+1}(1 - 2a_j^\dagger a_j)(1 - 2b_j^\dagger b_j)(1 - 2c_j^\dagger c_j) \prod_{j<l}(1 - 2d_j^\dagger d_j)(1 - 2e_j^\dagger e_j)$$

$$S_{e,l}^+ = e_l^\dagger \prod_{j<l+1}(1 - 2a_j^\dagger a_j)(1 - 2b_j^\dagger b_j)(1 - 2c_j^\dagger c_j)(1 - 2d_j^\dagger d_j) \prod_{j<l}(1 - 2e_j^\dagger e_j)$$

$$S_{m,l}^z = m_l^\dagger m_l - \frac{1}{2} \quad (m = a, b, c, d, e) \quad (5)$$

where *a*, *b*, *c*, *d*, and *e* are fermionic annihilation operators corresponding to the spin-1/2 sites within each unit cell. Using the Jordan–Wigner transformation, the spin interaction terms in the Hamiltonian can be expressed in terms of these fermionic operators. Specifically, a typical spin-flip term such as $S_{a,l}^+ S_{b,l}^-$ transforms to $a_l^\dagger b_l$, where $a_l^\dagger$ and $b_l$ are the fermionic creation and annihilation operators at sites *a* and *b* in the *l*-th unit cell, respectively. Consequently, the Hamiltonian can be reduced to a free spinless fermion model, which takes the following form:

$$H = \frac{1}{2}\sum_{l=1}^{N}\left\{\left[J_1(a_l^\dagger b_l + b_l^\dagger c_l) + J_2\{c_l^\dagger d_l + \frac{1}{2}(e_{l-1}^\dagger a_l + e_l^\dagger a_{l+1})\} + J_3(d_l^\dagger e_l)\right] + h.c. - 2h[(a_l^\dagger a_l - \frac{1}{2}) + (b_l^\dagger b_l - \frac{1}{2}) + (c_l^\dagger c_l - \frac{1}{2}) + (d_l^\dagger d_l - \frac{1}{2}) + (e_l^\dagger e_l - \frac{1}{2})]\right\} \quad (6)$$

The Fourier transformation of Fermi operator to momentum space (*k*-space) is defined as

$$m_l^\dagger = \frac{1}{\sqrt{N}}\sum_k e^{-ikl} m_k^\dagger \quad \text{with } (m = a,b,c,d,e) \quad (7)$$

Therefore, the Hamiltonian for the interacting trimer-dimer spin chain in the *k*- space, considering unit spacing between adjacent spins in a chain, can be written as

$$H = \frac{1}{2}\sum_k \left\{\left[J_1(a_k^\dagger b_k + b_k^\dagger c_k) + J_2(c_k^\dagger d_k + e^{ik} e_k^\dagger a_k) + J_3(d_k^\dagger e_k)\right] + h.c. - 2h(a_k^\dagger a_k + b_k^\dagger b_k + c_k^\dagger c_k + d_k^\dagger d_k + e_l^\dagger e_l) + \frac{5}{2}Nh\right\} \quad (8)$$

Here, the summation of the wavevector *k* spans over the first Brillouin zone, assuming periodic boundary conditions. Notably, string operators emerge only when exchange interaction $J_2$ spans between adjacent unit cells (*l*-1) and *l* (Fig. 1). In contrast, such terms cancel out when the interaction remains confined within the same unit cell. The string terms introduce dispersion in the excitation spectrum, making the analysis nontrivial. To investigate the nature of these excitations and the resulting magnetization behaviour, we employ Green's function formalism. The Green-function approach is widely used for the study of free-fermion spin models [49-54] as it provides a single unified object, $G(k,\omega)$, from which magnetization, correlation functions, excitation spectra, and thermodynamic quantities can be derived directly in a consistent manner. This leads to transparent analytical expressions and avoids observable-specific constructions. The Green's function composed of fermion spin operators may be defined as:

$$G_{k,k'} = \begin{pmatrix} \ll a_k; a_k^\dagger \gg & \ll a_k; b_k^\dagger \gg & \ll a_k; c_k^\dagger \gg & \ll a_k; d_k^\dagger \gg & \ll a_k; e_k^\dagger \gg \\ \ll b_k; a_k^\dagger \gg & \ll b_k; b_k^\dagger \gg & \ll b_k; c_k^\dagger \gg & \ll b_k; d_k^\dagger \gg & \ll b_k; e_k^\dagger \gg \\ \ll c_k; a_k^\dagger \gg & \ll c_k; b_k^\dagger \gg & \ll c_k; c_k^\dagger \gg & \ll c_k; d_k^\dagger \gg & \ll c_k; e_k^\dagger \gg \\ \ll d_k; a_k^\dagger \gg & \ll d_k; b_k^\dagger \gg & \ll d_k; c_k^\dagger \gg & \ll d_k; d_k^\dagger \gg & \ll d_k; e_k^\dagger \gg \\ \ll e; a_k^\dagger \gg & \ll a_k; b_k^\dagger \gg & \ll a_k; c_k^\dagger \gg & \ll a_k; d_k^\dagger \gg & \ll a_k; e_k^\dagger \gg \end{pmatrix}$$

(9)

When the Green's function is put into the equation of motion, we obtain the following matrix equation

$$M \cdot G_{k,k'} = \delta_{k,k'} \cdot I \quad (10)$$

where,

$$M = \begin{pmatrix} \omega+h & -\frac{J_1}{2} & 0 & 0 & -\frac{J_2}{2}e^{-ik} \\ -\frac{J_1}{2} & \omega+h & -\frac{J_1}{2} & 0 & 0 \\ 0 & -\frac{J_1}{2} & \omega+h & -\frac{J_2}{2} & 0 \\ 0 & 0 & -\frac{J_2}{2} & \omega+h & -\frac{J_3}{2} \\ -\frac{J_2}{2}e^{ik} & 0 & 0 & -\frac{J_3}{2} & \omega+h \end{pmatrix} \text{ and } I = \begin{pmatrix} 1 & 0 & 0 & 0 & 0 \\ 0 & 1 & 0 & 0 & 0 \\ 0 & 0 & 1 & 0 & 0 \\ 0 & 0 & 0 & 1 & 0 \\ 0 & 0 & 0 & 0 & 1 \end{pmatrix}$$

The elementary excitation spectra can be obtained from the above equation by imposing the condition

$$\det(M) = 0 \tag{11}$$

which yields a fifth-degree polynomial in $[\omega(k) + h]$. The roots of this characteristic polynomial, denoted as $R_i(k)$ (where $i = 1,2,\ldots 5$), correspond to the elementary quasiparticle excitations of the interacting trimer-dimer spin chain. However, according the Abel's theorem "there exists no general algebraic solution for polynomial equations of degree five or higher using radicals" [55]. As a result, analytical solution for obtaining elementary excitations is not possible in the present case. Therefore, we employed numerical methods to compute the excitation spectrum $R_i(k)$, and to explore the dispersion relations, band gaps, and the conditions for magnetization plateaus.

Further, to compute the magnetization as a function of magnetic field and temperature, we employed the standard spectral theorem. According to this theorem the correlation function involving the product of the fermionic operators can be evaluated using the corresponding Green's function:

$$< m_k^\dagger m_k > = \frac{i}{2\pi} \int_{-\infty}^{\infty} \frac{<< m_k; m_k^\dagger >>_{\omega+i0^+} - << m_k; m_k^\dagger >>_{\omega-i0^+}}{\exp(\beta\omega) + 1} d\omega, \quad m = a,b,c,d,e \tag{12}$$

here $\beta$ is the inverse temperature ($= 1/T$). The green function $<< m_k; m_k^\dagger >>$ can be obtained from the above matrix Eq. (9). For the further consideration, it is useful to remind that the down (up) spin corresponds empty (filled) fermion state. Thus, the sublattice magnetization can be obtained from the fermionic average in the following way [Ref]:

$$< S_{m,l}^z > = \sum_k < m_k^\dagger m_k > - \frac{1}{2} \; ; (m = a,b,c,d,e) \tag{13}$$

(13)

Therefore, the average reduced magnetization ($M$) per global unit (trimer-dimer) is defined as,

$$M = \frac{1}{5N} \sum_{p=a,b,c,d,e} \sum_{l=1}^{N} \langle S_{p,l}^z \rangle = \frac{1}{5N} \sum_{l=1}^{N} \langle (S_{a,l}^z + S_{b,l}^z + S_{c,l}^z + S_{d,l}^z + S_{e,l}^z) \rangle \tag{14}$$

$$M = -\frac{1}{2} \int_{-\infty}^{\infty} dE \, \rho(E) \tanh\left(\frac{\beta E}{2}\right) \tag{15}$$

where the density of state $\rho(E)$ is defined as

$$\rho(E) = \frac{1}{5N} \sum_{p=1}^{5} \sum_{i=1}^{N} \delta(E - E_{p,i}) \tag{16}$$

The above formalisms of the magnetization $M$ and excitation spectrum $\omega$ indicate that their natures depend on the exchange couplings $Js$ and the applied magnetic field $h$. In the following sections, we discuss in detail the behaviors of *M* and excitations of the spin-½ XX hybrid trimer–dimer chain as a function of *Js* and *h*. Based on the derived magnetization behaviours, we construct a ground-state phase diagram in the *h*–*J* plane.

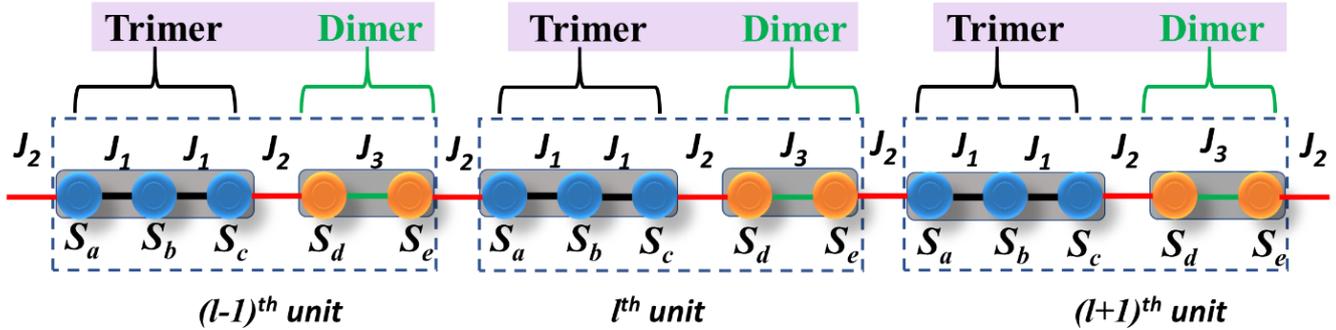

**Figure 1.** Schematic representation of the interacting trimer-dimer spin-1/2 chain with exchange interaction topology $J_1$-$J_1$-$J_2$-$J_3$. The circles denote spin-1/2 sites, which are coupled via exchange interactions $J_1$, $J_2$, and $J_3$. Here, $J_i < 0$ and $J_i > 0$ represent ferromagnetic (FM) and antiferromagnetic (AFM) couplings, respectively. The index *l* denotes the unit cell number, with each unit cell consisting of five spin sites or subcells labelled as $S_a$, $S_b$, $S_c$, $S_d$ and $S_e$.

### III. RESULTS AND DISCUSSION:

In general, the ground state of the quantum magnets is highly sensitive to the relative strengths of the exchange interactions within their Hamiltonian. In this section, we investigate the influence of the exchange parameters $J_1$, $J_2$, and $J_3$ on the magnetic properties of the spin-1/2 interacting trimer-dimer chain. Particular attention is given to the magnetization behavior under the application of a transverse magnetic field (*h*). We systematically analyze the conditions under which magnetization plateaus emerge or vanish as a function of the relative strengths of the intratrimer ($J_1$), inter-unit ($J_2$), and intradimer ($J_3$) exchange couplings. Additionally, the effect of temperature on the magnetization plateau states is investigated through detailed calculations of elementary excitations. These analyses provide comprehensive insights into the interplay between exchange interactions, magnetic field, and temperature in determining the magnetic properties of the system.

#### 1.2. Quantum Magnetization Plateau State

The magnetization curves of the spin-1/2 interacting trimer-dimer chain system have been calculated using Eq. (14) as a function of the external magnetic field (*h*) for various sets of exchange interactions $J_1$, $J_2$, and $J_3$. The influence of different interaction regimes on the magnetization process have been investigated. Specifically, we consider the following scenarios: (i) a dominant intratrimer exchange interaction $J_1$, (ii) a dominant intradimer exchange interaction $J_3$, (iii) a strong inter dimer-trimer exchange interaction $J_2$, and (iv) the uniform limit where all the exchange interactions are equal in strength ($J_1=J_2=J_3$). For all calculations and subsequent discussions, we have considered the same unit (Kelvin) for both the magnetic field (*h*) and exchange interactions $J_i$ as defined in Eq. 2.

Figure 2 shows the calculated zero-temperature magnetization curves for a fixed value of $J_2$ (=0.5) while systematically varying the values of $J_1$ and $J_3$. Figures 2 (a-c) show the magnetization profiles for a constant value of $J_1$ (= 1.0), with $J_3$ = 0.25, 0.5, and 0.625, respectively. For all cases, as the magnetic field increases, the magnetization initially rises and reaching to a value of $M_s/5$ at a field $h_1$. Further increasing in magnetic field, the magnetization remains constant up to a field $h_2$, revealing the formation of a 1/5 magnetization plateau state. Beyond $h_2$, the magnetization value increases again up to $3M_s/5$ at field $h_3$, and remain constant up to afield value of $h_4$, revealing a second plateau at 3/5 of the saturation magnetization.

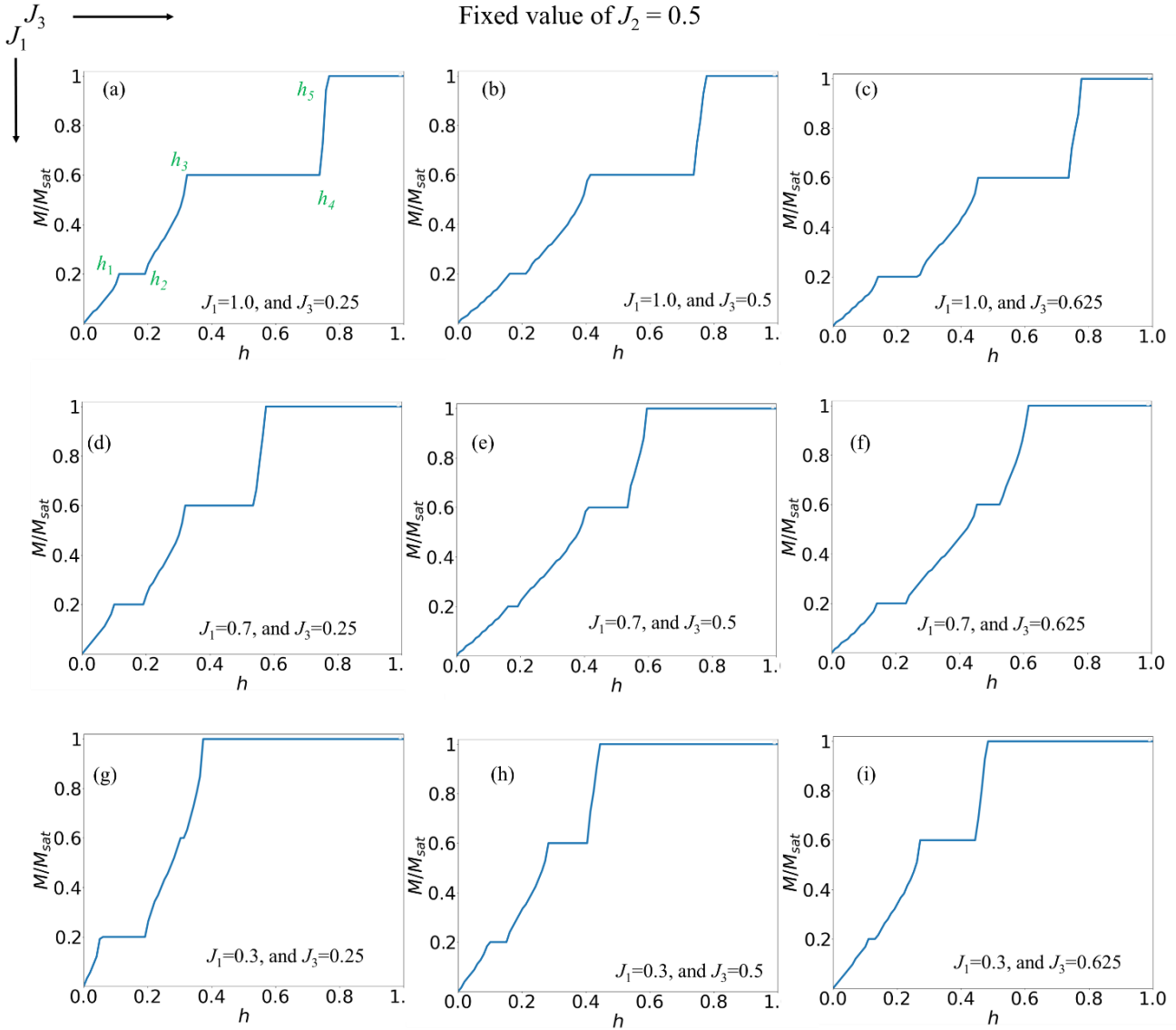

**Figure 2.** Magnetic field dependence of the magnetization ($M$ vs $h$) for the interacting trimer-dimer spin-1/2 chain for various combinations of intratrimer ($J_1$) and intradimer ($J_3$) exchange interactions, with a fixed value of inter trimer-dimer exchange interaction $J_2 = 0.5$. $M$ vs $h$ curves for (a-c) $J_1 =1.0$, with $J_3 = 0.25, 0.5$ and $0.625$, respectively; (d-f) $J_1 = 0.7$, with $J_3=0.25, 0.5, 0.625$, respectively; and (g-i) $J_1 = 0.3$, with $J_3 = 0.25, 0.5$ and $0.625$, respectively.

A further increase in the field beyond $h_4$ results in a continuous rise of magnetization until it reaches the full saturation value at a saturation field $h_5$. Thus, the presence of quantized magnetization plateaus at $M = M_s/5$ and $M = 3M_s/5$ is clearly evident at $T = 0$ K. These plateaus satisfy the Oshikawa–Yamanaka–Affleck (OYA) criterion stated in Equation (1), considering a periodicity of five spins per unit cell and $S = 1/2$. This confirms that the emergence of magnetization plateaus in the hybrid trimer-dimer spin chain originates from the global periodicity of the spin-chain, rather than the periodicity of the isolated clusters, i.e., trimer or dimer units.

The widths of the magnetization plateaus, defined by $(h_2-h_1)$ for the 1/5 plateau and by $(h_4-h_3)$ for the 3/5 plateau, depend on the intra-dimer exchange interaction $J_3$. For fixed values of $J_1=1.0$ and $J_2=0.5$, increasing $J_3$ from 0.25 to 0.625 leads to a non-monotonic variation (decrease up to $J_3 = 0.5$ and then increases) of 1/5 plateau width. In contrast, the width of the 3/5 plateau decreases monotonically with increasing $J_3$. Interestingly, the saturation field $h_5$ remains unchanged with $J_3$, indicating that the energy required to fully polarize the system is not affected by $J_3$ within this parameter regime. Figures 2 (d-f) show the magnetization curves for a reduced intra-trimer exchange interaction $J_1 = 0.7$, with varying $J_3 = 0.25$, 0.5, and 0.625. In this case, both 1/5 and 3/5 magnetization plateaus are still observed, and the dependences of their widths on $J_3$ follow the same qualitative trends as observed for $J_1=1.0$. However, the saturation field $h_5$ is found to be lower than that in the stronger $J_1$ regime, indicating that a weaker trimer coupling reduces the overall energy scale required to reach full magnetization. Figures 2(g-i) present the magnetization curves for further reduced value of $J_1=0.3$. For this case, a different trend emerges: the width of the 1/5 plateau decreases monotonically with increasing $J_3$, while the width of the 3/5 plateau increases monotonically. This contrast trend highlights the complex interplay between the intratrimer ($J_1$) and intradimer ($J_3$) exchange interactions in determining the stability of quantized magnetization states. Our comprehensive calculations, reveal that one of the magnetization plateaus vanishes entirely for a specific combination of $J_1$ and $J_3$. The parameter points at which such transitions occur are identified as quantum critical points in the $J_1$– $J_2$– $J_3$ parameter space. These critical points demarcate distinct quantum phases and will be discussed in detail in a subsequent section.

Figure 3 depicts the magnetization curves at three selected points in the $J_1$– $J_2$– $J_3$ parameter space. Figures 3(a) and 3(b) correspond to the parameters values $J_3=0.7$ and $J_3=0.2$ with $J_1=0.3$, $J_2=0.5$, respectively. The magnetization curve shows the absence of the 1/5 magnetization plateau for $J_3=0.7$ [Fig. 3(a)] and the absence of the 3/5 magnetization plateau for $J_3=0.2$ [Fig. 3(b)]. These determine the presence of quantum critical points at (i) $J_1=0.3$, $J_2=0.5$ and $J_3=0.7$ and (ii) $J_1=0.3$, $J_2=0.5$ and $J_3=0.2$, where one of the plateau states vanishes. Figure 3(c) shows the magnetization behaviour for the uniform case, $J_1= J_2= J_3= 1.0$, in which no magnetization plateau is observed. The smooth and continuous magnetization curve in this case is consistent with the behavior expected for a uniform spin-1/2 XX chain, as originally described by Griffiths *et al.* [56]. These results clearly demonstrate that both the number of magnetization plateau states and their characteristic width are highly sensitive to the relative strengths of the exchange interactions $J_1$, $J_2$, and $J_3$. The variations of critical magnetic fields $h_1$, $h_2$, $h_3$, and $h_4$ as a function of $J_3$ is shown in Figs. 4(a-d). While the fields $h_3$, and $h_4$ vary monotonically with $J_3$, the fields $h_1$ and $h_2$, show non-monotonic behaviors. In contrast, the saturation field $h_5$ (not shown here) monotonically decreases with increasing $J_3$ regardless of the values of $J_2/J_1$.

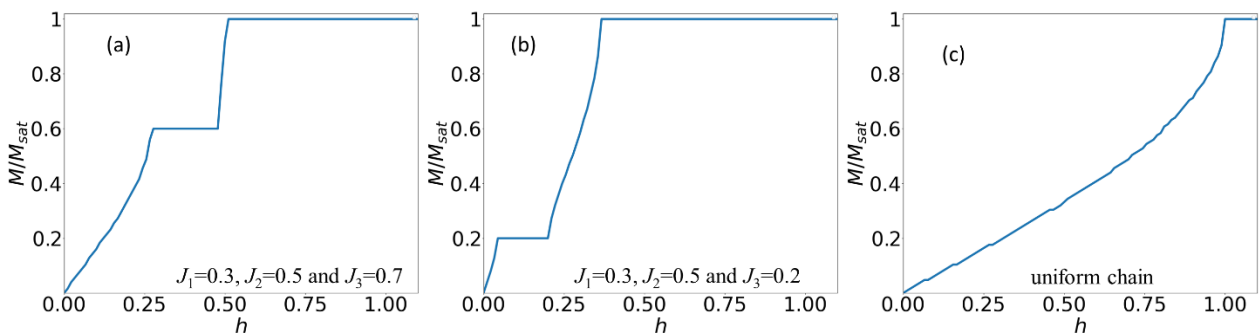

**Figure 3.** Field dependence magnetization curves of the interacting trimer-dimer spin-1/2 chain for selected combinations of $J_1$, $J_2$, and $J_3$ exchange interactions: (a) $J_1=0.3$, $J_2=0.5$ and $J_3=0.7$, where the 1/5 plateau is absent; (b) $J_1=0.3$, $J_2=0.5$ and $J_3=0.2$, where the 3/5 plateau is absent; and (c) uniform case $J_1= J_2= J_3=1.0$, where both the 1/5 and 3/5 plateaus are absent.

The widths of the 1/5 and 3/5 magnetization plateaus as a function of $J_3$ are shown in Figs. 5(a) and 5(b) for different values of $J_2/J_1$, respectively. It is evident that the 1/5 and 3/5 magnetization plateau states are present across most of the ($J_1$, $J_2$ and $J_3$) exchange interaction parameter space except specific combinations of these parameters. For the 1/5 magnetization plateau, for a stronger intra-trimer coupling $J_1 = 2$, the plateau vanishes for a weaker intra-dimer coupling $J_3 = 0.2$. As the strength of intra-trimer coupling $J_1$ decreases from $J_1 =2$ to $J_1 =0.3$, the critical value of intra-dimer coupling $J_3$ increases from $J_3 =0.2$ to $J_3 =0.8$, respectively. On the other hand, for the 3/5 magnetization plateau, for a stronger intra-trimer coupling $J_1 = 2$, the plateau disappears for a stronger intra-dimer coupling $J_3 >1$. As the strength of intra-trimer coupling $J_1$ decreases, the critical value of intra-dimer coupling $J_3$ also decreases which is opposite to the behaviour of the 1/5 magnetization plateau.

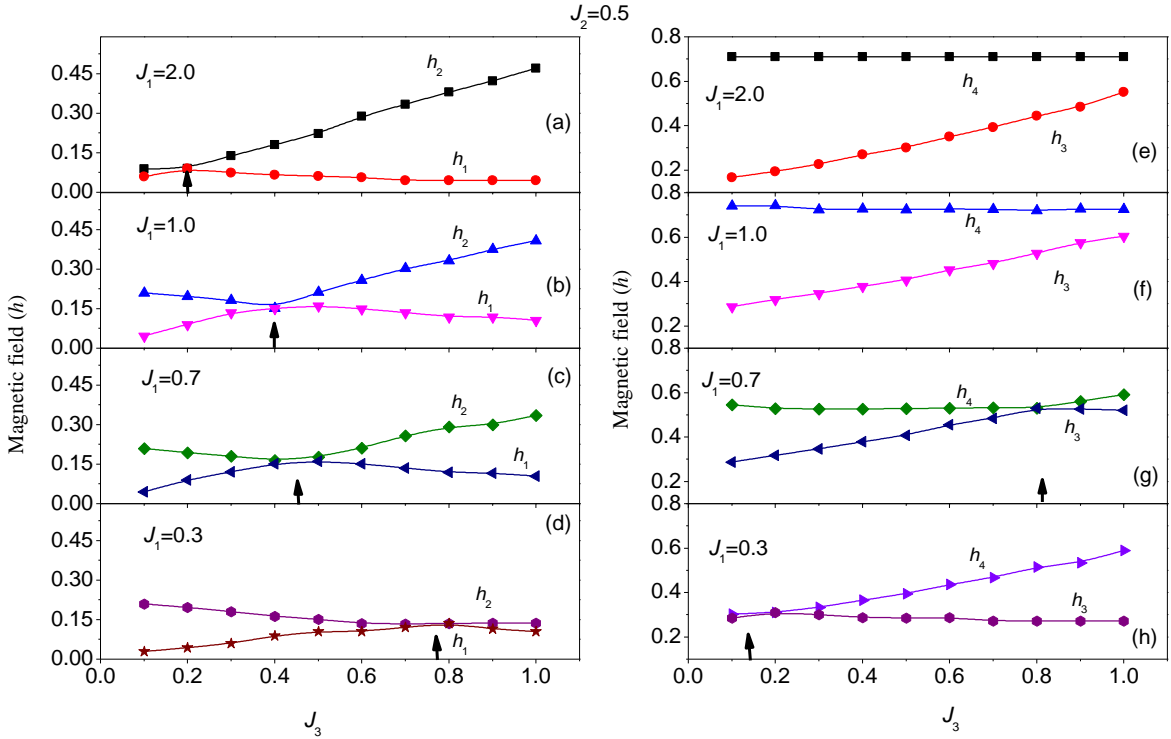

**Figure 4.** Variation of the critical fields (a-d) $h_1$ and $h_2$, and (e-f) $h_3$ and $h_4$ as a function of $J_3$ for different values of $J_1$ and a fixed value of $J_2 = 0.5$. The vertical arrow indicates the critical values of $J_3$ at which magnetization plateau vanishes.

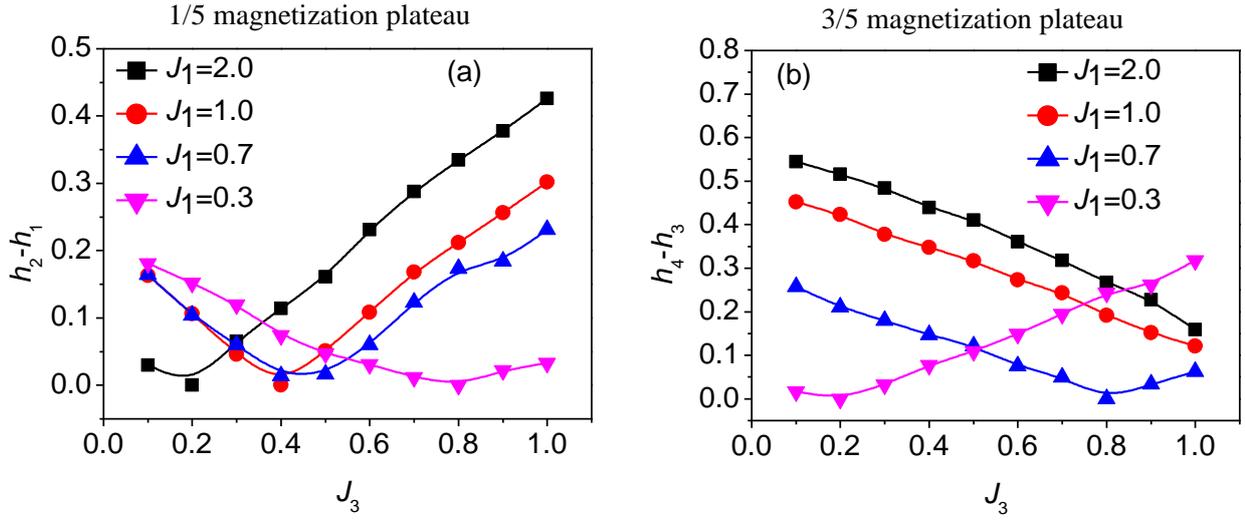

**Figure 5.** The variation of the widths of the (a) 1/5 and (b) 3/5 magnetization plateaus of hybrid trimer-dimer spin-1/2 XX chain as a function of $J_3$ and $J_1$ with a fixed value of $J_2 = 0.5$.

### 1.3. Elementary Excitations and Phase Diagram

To understand the origin of the 1/5 and 3/5 magnetization plateaus in the interacting hybrid trimer-dimer spin-1/2 chain system, we have calculated the magnetic excitation spectra [Fig. 6] for different sets of exchange interactions using the model Hamiltonian [Eq. (2)]. The Fermi level of the JW fermions is set to zero *i.e.* $E_F = 0$. Since, the spin system has been mapped to spinless fermionic particles through the JW transformation, the magnetic field $h$ acts as the chemical potential of the fermions, and the magnetization $M$ corresponds to the number of fermions per unit cell occupying the states below the Fermi level. The excitation spectrum consists of five energy bands, obtained from the solution of Eq. (11). For $J_1=1.0$, $J_2=0.5$, and $J_3=0.25$, one energy band crosses the Fermi level at zero field, two bands lying above the $E_F$, and two bands lying below the $E_F$. The ground state at $h = 0$ corresponds to a configuration in which all states with $E(k) < E_F$ are occupied and while those with $E(k) > E_F$ remains empty. Upon applying an external magnetic field, all energy bands shift downward, which is equivalent to raising the effective Fermi level. At $h = 0$, the intermediate-energy band $E_3(k)$ crosses the Fermi level at two momentum points located at $k_F = \pm\pi/2$ within the first Brillouin zone ($-\pi < k < \pi$), resulting in net zero magnetization. As the field increases, this band gradually fills between $\pm\pi/2$ and $\pm\pi$, contributing to a linear rise in magnetization until the Fermi level reaches the maximum point of the band at $k = \pm\pi$ for a field $h_1$.

In the magnetic field range ($0 < h < h_1$), the excitation is gapless and the magnetic state corresponds to an exotic quantum phase known as Luttinger liquid (LL) phase, commonly observed in one-dimension spin systems [57-59]. As magnetic field increases above $h_1$, the Fermi level crosses the top of the intermediate-energy band $E_3(k)$, and enters in the region of energy gap between $E_3(k)$ and $E_4(k)$ bands. In this regime, the number of occupied states below the Fermi level remains constant, leading to a magnetization plateau at 1/5 of the saturation magnetization. The magnetization value remains constant at $M_s/5$ until a magnetic field of $h_2$ where Fermi level touches the bottom of the $E_4(k)$ band. The width of the 1/5 plateau ($h_2-h_1$) corresponds to the energy gap $g_1$ between $E_3(k)$ and $E_4(k)$ bands at $k=\pm\pi$. Beyond $h_2$, the Fermi level enters into the $E_4(k)$ band, and the magnetization rises again as number of the occupied states below the Fermi level increases. This corresponds to another gapless region (Phase-III). The magnetization value continues to increase until the Fermi level reaches the top point of the $E_4(k)$ band at $h_3$ and reaches to 3/5 of the saturation magnetization. The 3/5 magnetization plateau is observed when the Fermi level lies within the energy gap between the $E_4(k)$ and $E_5(k)$ bands, for applied field between $h_3$ and $h_4$. The plateau width ($h_4-h_3$) is proportional to the energy gap value $g_2$ between these two bands. As the field increases beyond $h_4$, the Fermi level enters the band $E_5(k)$, giving rise to another gapless region (Phase-V). The system reaches full saturation when the Fermi level exceeds the top of the $E_5(k)$ band. It is important to note that the widths of the magnetization plateaus are directly proportional to the energy gaps between the successive energy bands in the excitation spectrum. While, the widths of the gapless phases (LL, Phase- III and Phase-V) are determined by the bandwidths ($\gamma$) of the individual energy bands. The bandwidth $\gamma$ is defined as the energy difference between the zone centre ($k = 0$) and zone boundary ($k= \pi$) i.e. $\gamma = |E(k=0) - E(k=\pi)|$. Table-I summarizes the ranges of magnetic field and corresponding positions of the Fermi level in the excitation spectrum for the distinct magnetic phases (Phase-I to Phase-VI).

**Table-I** Magnetic field ranges corresponding to different magnetic phases and their associated Fermi level positions in the elementary excitation spectrum.

| Phase | Magnetic field range | Position of Fermi level |
|---|---|---|
| Phase-I: LL Phase | 0- $h_1$ | $E_3(k= \pi/2)$ to $E_3(k= \pi)$ |
| Phase-II: 1/5 plateau | $h_1$-$h_2$ | $E_3(k= \pi)$ to $E_4(k= \pi)$ |
| Phase-III: Gapless phase-I | $h_2$-$h_3$ | $E_4(k= \pi)$ to $E_4(k= 0)$ |
| Phase-IV: 3/5 plateau | $h_3$-$h_4$ | $E_4(k=0)$ to $E_5(k=0)$ |
| Phase-V: Gapless phase-II | $h_4$-$h_5$ | $E_5(k=0)$ to $E_5(k= \pi)$ |
| Phase-VI: Saturation/Fully polarized phase | beyond $h_5$ | above $E_5(k= \pi)$ |

Our calculations further reveal that energy gaps $g_1$ and $g_2$ vanish at the critical points where the 1/5 ($J_1$=0.3, $J_2$=0.5, and $J_3$=0.7) and 3/5 ($J_1$=0.3, $J_2$=0.5, and $J_3$=0.2) magnetization plateau, respectively disappear [Fig. 6(b-c)]. Furthermore, in the isotropic case with $J_1$=$J_2$=$J_3$=1.0, where both the 1/5 and 3/5 magnetization plateaus disappear [Fig. 3(c)], both energy gaps $g_1$ and $g_2$ become zero [Fig. 6(d)]. The parameter points (various combinations of $J_1$ and $J_3$) where the excitation spectrum becomes gapless and no magnetization plateau is observed, are identified as quantum critical points, denoted by "C" in the $h$-$J_2$ plane phase diagram [Fig. 7]. These results firmly establish that the magnetization plateaus are intimately connected to the inter-band energy gaps in the excitation spectrum. Notably, the 1/5 magnetization plateau is observed to be more robust against the exchange interactions, whereas, the 3/5 magnetization plateau is comparatively more sensitive and less stable.

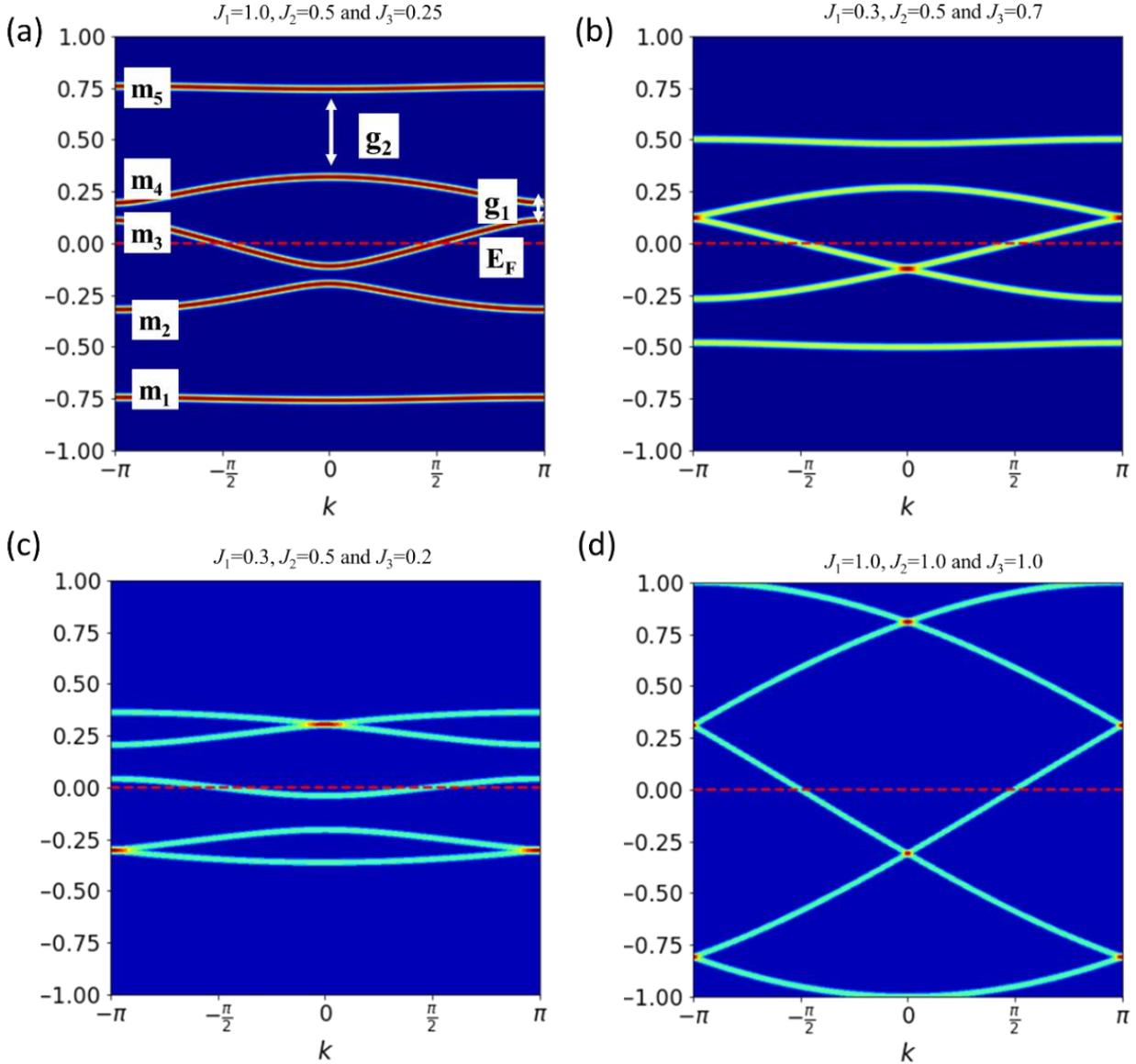

**Figure 6.** Elementary excitation spectra of the spin-1/2 trimer-dimer chain system for various sets of exchange interactions: (a) $J_1=1.0$, $J_2=0.5$ and $J_3=0.25$, (b) $J_1=0.3$, $J_2=0.5$ and $J_3=0.7$ (c) $J_1=0.3$, $J_2=0.5$ and $J_3=0.2$, and for (d) uniform XX chain with $J_1=1.0$, $J_2=1.0$ and $J_3=1.0$. The dashed line represents the Fermi level of Jordan–Wigner fermions. The five excitation modes are labeled as $m_1$ to $m_5$. The energy gaps $g_1$ and $g_2$ denote the inter-band gaps between $E_3(k)$ and $E_4(k)$, and between $E_4(k)$ and $E_5(k)$, respectively.

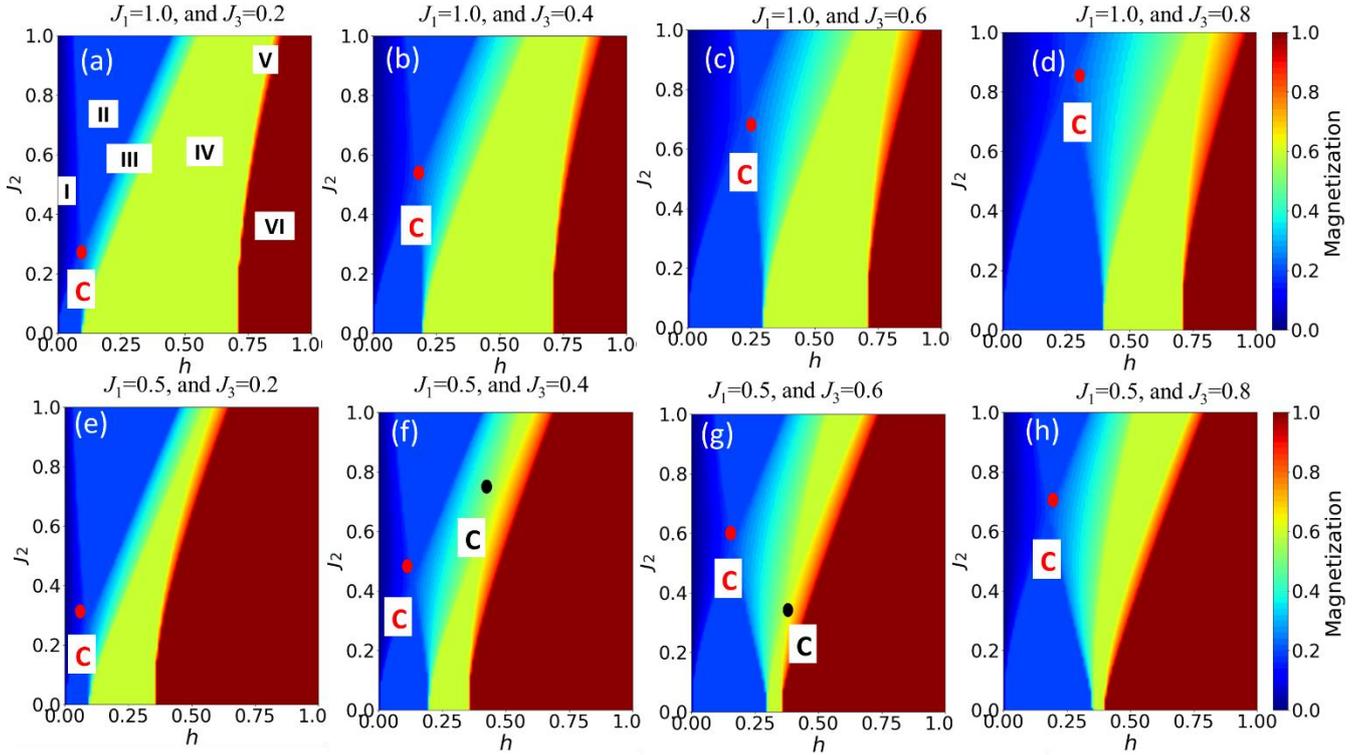

**Figure 7.** Phase diagrams in the ($h$-$J_2$) plane for spin-1/2 interacting trimer-dimer chain, shown for different sets of exchange interactions (a-d) fixed $J_1$=1.0 and $J_3$ = 0.2, 0.4, 0.6 and 0.8, (e-h) fixed $J_1$=0.5 and $J_3$ = 0.2, 0.4, 0.6 and 0.8. Distinct magnetic phases are: I. Luttinger liquid (LL) phase; II. 1/5 magnetization plateau phase, III. gapless phase, IV. 3/5 magnetization plateau phase, V. gapless phase, and VI. saturation phase. The point "C" denotes the quantum critical point at which the 1/5 (red) and 3/5 (black) magnetization plateaus vanish.

### 1.4. Temperature effect on the magnetization plateau states

Figure 8(a) shows the magnetization curves as a function of magnetic field at various temperatures for the exchange parameters $J_1$=1.7, $J_2$=0.1 and $J_3$=0.3. As temperature increases, the magnetization plateaus progressively become narrower, and eventually vanish beyond a critical temperature. The critical temperature depends on the relative strength of the exchange interactions, and consequently, on the width of the magnetization plateau. Plateau with smaller width melts at lower temperature. With increasing temperature, the slopes of the magnetization curves on both sides of the plateaus change symmetrically, implying a comparable density of states on either side of the plateau. Similar thermal melting of magnetization plateaus was reported in other low-D systems, viz., trimer-chain, Kagome, and triangular spin system[28,60,61].

Figure 8(b) illustrates the temperature dependence of magnetization curves under several fixed magnetic fields. The $M(T)$ curves exhibit distinct behavior at low temperatures depending on the applied magnetic field. In the field range $h_1$- $h_2$ (1/5 plateau range), the magnetization remains constant up to a critical temperature ($T^*$) and then show temperature dependence for higher temperatures. Above the $T^*$, the magnetization decreases monotonically with increasing temperature for the applied magnetic field corresponding to the first half of the 1/5 plateau i.e., $h_1 < h < (h_1 + h_2)/2$. Whereas, for the applied magnetic field corresponding to the second half of the 1/5 plateau i.e., $(h_1 + h_2)/2 < h < h_2$, the magnetization initially increases with increasing temperature above $T^*$, then decreases at further higher temperatures. The initial increase in the magnetization attributes to thermally activated occupation of states above the Fermi level.

A similar temperature dependence behaviour of the magnetization is observed near the 3/5 magnetization plateau. These observations suggest that the temperature and magnetic field-dependent behaviors arise from the different magnetic phases including the LL phase, gapped plateau states, and gapless regions. Notably, the magnetization curves show an inflection point at the centre of the 1/5 and 3/5 plateau regions, corresponding to field values $(h_1 + h_2)/2$ and $(h_3 + h_4)/2$, respectively. These fields may be identified as crossover points between distinct quantum phases in the spin system.

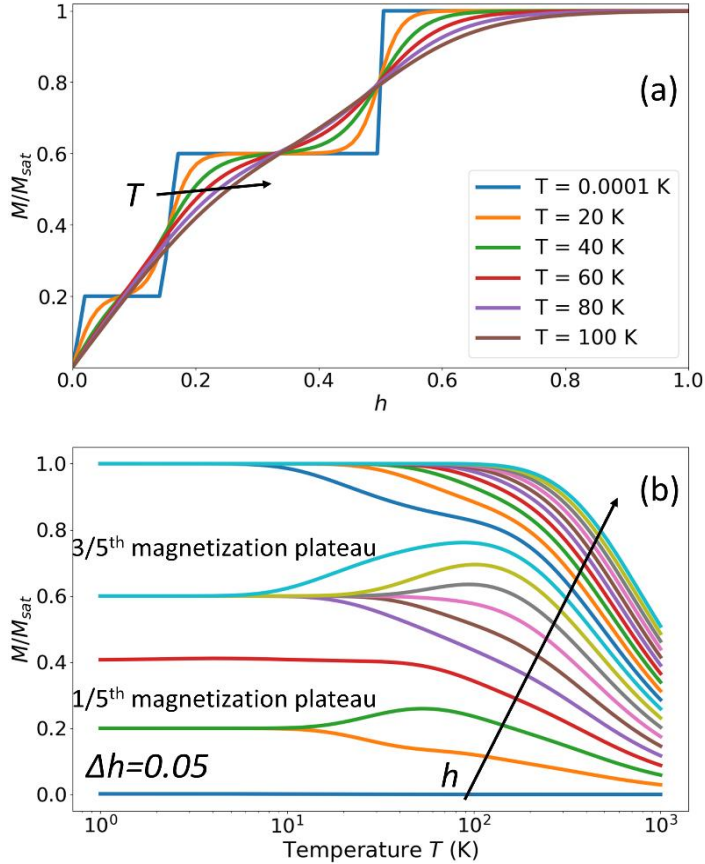

**Figure 8.** (a) Calculated magnetization curves as a function of the applied magnetic field $h$ at different temperatures, for the spin-1/2 trimer-dimer chain with exchange interactions $J_1=0.7$, $J_2=0.1$ and $J_3=0.3$. (b) The logarithmic temperature dependence of magnetization under various applied fields for the same set of exchange parameters, highlighting the contrasting thermal response across different field regimes.

### 1.5. Discussion:

In this work, we have shown that the emergence of magnetization plateaus in the hybrid trimer-dimer spin-1/2 chain can be traced directly to the interplay between exchange couplings ($J_1$: intratrimer exchange interaction, $J_2$: inter trimer-dimer exchange interaction, and $J_3$: intradimer exchange interaction) and the resulting excitation spectrum. When $J_1$ dominates, the system favors strong trimer correlations with 1/3 magnetization plateau. On the other hand, for a dominant $J_3$, the system becomes a Spin ½ dimer without a plateau state at fractional value of the $M$s. The exchange interaction $J_2$ acts as a coupling channel that balances these competing tendencies. The 1/5 and 3/5 plateau states appear only when the competing exchange interactions stabilize gapped excitation states, with the width of each plateau controlled by the size of the interband excitation gap value $g_1$ and $g_2$. Closing of these gaps in the excitation spectra drives into gapless regimes through a quantum phase transition from gapped to gapless state, reflecting the intimate link between microscopic interactions and macroscopic magnetization steps. The different robustness of the two plateau states (where the 1/5 plateau survives over a broader interaction range than the 3/5 plateau) illustrates how the global periodicity of the hybrid trimer-dimer chain selects energetically favorable magnetic state. Thermal fluctuations further reshape the energy landscape and result into melting of plateau state above a characteristic temperature, giving rise to crossover features in the magnetization curves that distinguish gapped from gapless regimes. These findings conclude that the magnetization plateaus in hybrid trimer–dimer spin chains are not originating from the static spin configurations of the ground state, rather from spin-dynamic excitation spectrum. Our findings also shed light on the sensitivity of the plateau states to both internal exchange interactions and external perturbation like temperature.

Now, we detail below the nature of the magnetization plateaus for spin-1/2 chains composed solely either trimer or dimer units. For isolated spin-½ trimer chains, the characteristic magnetization plateau occurs at $\frac{M}{M_s} = 1/3$, which arises from the effective block-spin structure of the trimer and is consistent with the Oshikawa–Yamanaka–Affleck (OYA) condition. In contrast, a dimer chain yields only the trivial $\frac{M}{M_s} = 0$ and $\frac{M}{M_s} = 1$, corresponding to the singlet ground state and the fully polarized state, unless additional frustration is present. For an interacting trimer–dimer spin-1/2 chain, the overall periodicity becomes $Q = 3+2 = 5$. According to the OYA condition with $Q = 5$, the allowed magnetization plateaus are $\frac{M}{M_s} = 1/5$, 3/5, and 1 consistent with our results. This concludes that the magnetization plateau states are governed by the global periodicity of the chain ($Q = 5$), rather than the local periodicity of individual units ($Q = 3$ for a trimer or $Q = 2$ for a dimer), even when the intercluster exchange $J_2$ is weak [Fig 7]. In other words, the emergence of fractional plateau states and other properties is a direct manifestation of how exchange interactions enforce the global topology/periodicity of the spin Hamiltonian. With the similar analogy, other hybrid cluster spin-1/2 chain system, such as, tetramer–trimer with $Q = 7$, should exhibit fractionalized magnetization plateau at $\frac{M}{M_s} = 1/7$, 3/7, and 5/7, which demands experimental verifications on model compounds.

Although the present work focuses on the purely isotropic spin–½ XX trimer–dimer Hamiltonian, we realize that real spin-1/2 (*i.e.* $Cu^{2+}$) based compounds are more accurately described by Heisenberg (XXZ-type) interactions. At the XX limit, employed here, the Hamiltonian is exactly solvable as the Jordan–Wigner transformation maps it onto a free-fermion problem. In contrast, adding the longitudinal interaction $S_{a,l}^z S_{b,l}^z = (a_l^\dagger a_l - \frac{1}{2})(b_l^\dagger b_l - \frac{1}{2})$ in the Hamiltonian (Eq.2) introduces a quartic (density–density) fermionic term and makes the model interacting, which is beyond the free-fermion integrability. In this situation, the origin of magnetization plateaus becomes genuinely many-body. In this case, the $S_zS_z$ coupling gives interaction-induced self-energy corrections in the fermionic bands, and modifies the low-energy dispersion, that can open a spin gap or enhance the existing spin gap values. As a result, the critical fields are no longer determined solely by single-particle band-edge crossings—as in the pure XX case—and the plateau widths become renormalized. As reported for uniform spin-1/2 XXZ chains by Takahiro et al. [62](using the density matrix renormalization group method and a Jordan–Wigner fermion mean-field approximation), an increase of Ising-like $S_zS_z$ anisotropy shifts both the lower and upper critical fields to higher values with an overall expansion of the width of magnetization plateau, while the essential features of the plateau state remain intact. This behavior originates from a renormalization of the excitation spectrum with an effective increase in the excitation gap. Similar results, i.e., the broadening of the magnetization plateau with the increasing Ising-like anisotropy, are also obtained by bosonization approach, reported by Inoue et al. [63]. Moreover, for the frustrated $J_1$-$J_2$ spin-½ XXZ chain as well, Verkholyak et al. [42] showed (using a Jordan–Wigner mean-field analysis and exact diagonalization) that an enhancement of Ising-like anisotropy increase the excitation gap, hence, a broadening of the magnetization plateau width. Verkholyak et al. [42] also reported that in the lower limit of $J_2/J_1$, where the excitations are gapless, the increase in Ising-like anisotropy could induce a gap in the excitation spectra. Likewise, in the present trimer–dimer system, with the increase of Ising anisotropy, an increase of the excitation gap is expected through an interaction-driven renormalization of the excitation spectrum. Therefore, critical fields are expected to shift to higher values with an increase in the magnetization plateau width. Such changes may lead to an overall reshaping of the h-J phase diagram. Nevertheless, the exactly solvable XX trimer–dimer model provides a valuable reference point, establishing the baseline for hybrid trimer–dimer spin systems against which interaction effects can be systematically compared. We note that incorporation of the full XXZ interactions in the present study requires numerical methods—such as self-consistent mean-field approaches or density-matrix renormalization group (DMRG) techniques—capable of treating interacting Jordan–Wigner fermions.

## IV. SUMMERY AND CONCLUSION:

In this work, we have comprehensively investigated the magnetic properties of a $S=1/2$ XX hybrid trimer-dimer cluster quantum spin-1/2 chain using Green's function theory combined with the Jordan-Wigner (JW) transformation. Our analysis reveals the emergence of 1/5 and 3/5 magnetization plateaus, which comply with the Oshikawa-Yamanaka-Affleck (OYA) condition with periodicity $Q = 5$. In extension to the OYA condition, the present study establishes generalized condition for magnetization plateaus for hybrid cluster chain systems, showing that the allowed magnetization plateau states are governed by the global periodicity of the chain, instead of the local periodicity of individual cluster units. The ground state phase diagram in the magnetic field ($h$)−exchange interaction ($J$) plane, constructed for various combinations of intratrimer ($J_1$), inter trimer-dimer ($J_2$) and intradimer ($J_3$) exchange interactions, uncover a rich variety of quantum phases. These includes the Luttinger liquid (LL) phase, 1/5 and 3/5 the magnetization plateaus, and two distinct gapless phases. Temperature-dependent studies reveal that magnetization plateaus melt by thermal fluctuations. Narrower plateau vanishes at lower temperatures as compared to the plateaus having broader width. The temperature evolution of magnetization curves under various magnetic fields displays characteristic behaviours across different quantum phases. Our calculated elementary excitation spectra provide insights into the magnetization process and show that the plateau widths are governed by the inter-band energy gaps, while the widths of the gapless regions are correlated with the bandwidth ($\gamma$) of respective excitation bands. Furthermore, we demonstrate that the stability and existence of magnetization plateaus are highly sensitive to the exchange interaction parameters. The hybrid trimer-dimer spin-chain exhibits distinct features as compared to spin-chains composed of either trimers or dimers alone, particularly in terms of the number and nature of magnetization plateaus and the nature of elementary excitations. We believe that this comprehensive study will stimulate further experimental and theoretical exploration of quantum spin properties of trimer-dimer spin-1/2 chain compounds. In particular, our findings pave the way for future investigations of the thermodynamic properties and excitation dynamics using other advance numerical techniques, such as quantum Monte Carlo, and density matrix renormalization group (DMRG) methods.

**Acknowledgments:**

SMY acknowledges the financial assistance from ANRF, DST, Govt. of India, under the J.C. Bose fellowship program (JCB/2023/000014).

*****************************************************************